\documentclass[aps,twocolumn,superscriptaddress,showpacs,showkeys]{revtex4}

\usepackage{graphics,graphicx,dcolumn,bm,fleqn,epic,eepic,float}
\usepackage{amssymb,amsmath,multirow,rotate,color,float}
\usepackage{times}

\begin{document}
\title{Coherence in scale-free networks of chaotic maps}
\date{\today}
\author{Pedro G.~\surname{Lind}}
\homepage[URL: ]{http://www.ica1.uni-stuttgart.de/~lind}
\affiliation{Institute for Computational Physics, 
             Universit\"at Stuttgart, Pfaffenwaldring 27, 
             D-70569 Stuttgart, Germany}
\affiliation{Centro de F\'{\i}sica Te\'orica e Computacional, 
             Av.~Prof.~Gama Pinto 2,
             1649-003 Lisbon, Portugal}
\author{Jason A.C.~\surname{Gallas}}
\homepage[URL: ]{http://www.ica1.uni-stuttgart.de/~jgallas}
\affiliation{Institute for Computational Physics, 
             Universit\"at Stuttgart, Pfaffenwaldring 27, 
             D-70569 Stuttgart, Germany}
\affiliation{Instituto de F\'\i sica, Universidade Federal do Rio Grande
             do Sul, 91501-970 Porto Alegre, Brazil}
\author{Hans J.~\surname{Herrmann}}
\homepage[URL: ]{http://www.ica1.uni-stuttgart.de/~hans}
\affiliation{Institute for Computational Physics,  
             Universit\"at Stuttgart, Pfaffenwaldring 27, 
             D-70569 Stuttgart, Germany}
\affiliation{Departamento de F\'{\i}sica, Universidade Federal do
             Cear\'a, 60451-970 Fortaleza, Brazil}

\begin{abstract}
We study fully synchronized states in scale-free networks of chaotic
logistic maps as a function of both dynamical and topological
parameters.  
Three different network topologies are considered:
(i) random scale-free topology, 
(ii) deterministic pseudo-fractal scale-free network, and 
(iii) Apollonian network.
For the random scale-free topology we find a coupling strength
threshold beyond which full synchronization is attained. 
This threshold scales as $k^{-\mu}$, where $k$ is the outgoing
connectivity and $\mu$ depends on the local nonlinearity.
For deterministic scale-free networks coherence is observed only 
when the coupling strength is proportional to the neighbor
connectivity. 
We show that the transition to coherence is of first-order 
and study the role of the most connected nodes in the collective
dynamics of oscillators in scale-free networks. 
\end{abstract}

\maketitle

\section{Introduction and model}
\label{sec:intro}

Recently, intensive research on the structure and dynamics of networks
has provided insight for many systems where they arise naturally
\cite{livro,albert02,dorogovtsevrev}. 
Complex networks appear in a wide variety of fields ranging from
lasers \cite{meucci02},
granular media \cite{snoeijer04,otto03},
quantum transport \cite{texier04},
colloidal suspensions \cite{tanaka04}, 
electrical circuits \cite{almond04}, and
time series analysis \cite{small02},
to heart rhythms \cite{stewart04},
epidemics \cite{moreno04,dezso02},
protein folding \cite{compiani04}, and
locomotion \cite{zhaoping04} among others
\cite{livro,albert02,dorogovtsevrev}.

From the mathematical point of view, a network is a graph, composed by
nodes or vertices and by their connections or edges \cite{albert02}.
When studying network dynamics one frequently assumes a regular structure 
where each node evolves according to some more or less
complicated dynamics, typically fixed points \cite{lind04}, limit
cycles \cite{strogatz00} or chaotic attractors \cite{pecora97,anteneodo04}. 
When studying network structure, one usually
neglects node dynamics and all complexity is introduced
by the way nodes are connected to each other, i.e.~by
the network topology. 
With respect to their topology, networks are usually divided into
three large classes \cite{albert02}: 
random networks, where all the nodes are randomly connected \cite{erdos},
small-world networks introduced recently by Watts and
Strogatz \cite{strogatz01,watts98} as a middle ground between regular
and random networks, and the scale-free networks (e.g.~Barab\'asi and
Albert \cite{barabasi99}), where growth and preferential attachment
are considered.

The next logical step toward real network dynamics is
to consider simultaneously structural and dynamic complexity.
One important question addressed in this context is to know if 
synchronization between oscillators in such complex topologies would
appear and under which conditions it prevails.
In fact, coherent behavior of oscillator networks with complex
topologies has been studied for the random topology
\cite{manrubia99,jost01} and small-world topology
\cite{nishikawa03,barahona02,lago-fernandez00,hong04}. 
However, apart from a few exceptions \cite{jost01,jostalso,ieee}, 
there is a quite general lack of studies tackling
synchronization of chaotic oscillators in scale-free topologies.

In this paper we present detailed results concerning synchronization in
oscillator networks with scale-free topologies.
Our purpose is to determine under which conditions scale-free topologies
enable the emergence of coherent behavior.
As a general result, we present evidences that the transition to
synchronization is of first-order.
Our model reads
\begin{equation}
x_{t+1,i} = (1-\varepsilon)f(x_{t,i})+\frac{\varepsilon}{{\cal N}_i}
             \sum_{j\in {\cal K}_i} k_j^{\alpha}f(x_{t,j}) ,
\label{model}
\end{equation}
where $i=1,\dots,L$ and $t$ label discrete space and time respectively,
$L$ being the total number of oscillators,
$0 \le \varepsilon\le 1$ is the coupling parameter,
${\cal K}_i$ represents the set of labels of the neighbors of
node $i$, $k_i$ represents the number of such neighbors, 
and ${\cal N}_i=\sum_{j\in {\cal K}_i} k_j^{\alpha}$ normalizes the
interaction term in Eq.~(\ref{model}).
The function $f(x)$ is a continuous function governing node dynamics 
when connections are absent.
Here we choose the well-known quadratic map $f(x)=1-ax^2$, where
the free parameter $a$ is restricted to the interval
$-0.25\le a\le 2$ and contains all possible dynamical regimes from a
fixed point (e.g.~a=0) to fully developed chaotic orbits (e.g.~$a=2$).
The parameter $\alpha$ is a real number controlling the homogeneity in
the coupling: positive values of $\alpha$ enhance the coupling
strength with sites having larger number of neighbors, while
negative values favor sites having less neighbors. 
For $\alpha=0$ the coupling between each site and
its neighborhood is homogeneous, i.e.~it is independent on the
coordination. 

The linear stability of the coherent states $x_{t,i}=X, \forall i$ 
is governed by the variational equations of Eq.~(\ref{model}), whose diagonal
form reads \cite{pecora98,fink00,manrubiabook}
\begin{eqnarray}
\xi_{t+1,i}&=&\exp{\left (\Lambda (\varepsilon\lambda_i)\right )}\xi_{t,i}\cr
           & & \cr
           &=&\left [ Df(X)-\varepsilon\lambda_i Df(X)  \right ]\xi_{t,i},
\label{linestab}
\end{eqnarray}
where $\Lambda(\varepsilon\lambda_i)$ is the Lyapunov exponent, 
$Df(X)$ represents the identity matrix multiplied by the derivative of 
$f(x)$ computed at $x=X$ and $\lambda_i$ are eigenvalues of the coupling 
matrix $\mathbb{G}$ whose diagonal values are $g_{ii}=1$, 
while off-diagonal elements are
$g_{ij}=-k_j^{\alpha}/{\cal N}_i$ if nodes $i$ and $j$ are coupled 
and zero otherwise.
If $\mathbb{G}$ has zero-sum rows and all its eigenvalues 
$\lambda_1\le\lambda_2\le\dots\le\lambda_L$ are real,
then $\lambda_1=0$ corresponds to the mode parallel to the 
synchronization manifold and the largest Lyapunov exponent 
defines a master stability function \cite{pecora98}. 
The coherent state is stable whenever $\Lambda(\varepsilon\lambda_i)<0$
for $i=2,\dots,L$ \cite{pecora98,fink00,manrubiabook}.
In our case, it is easy to check that indeed $\mathbb{G}$ have zero-row sum,
yielding $\lambda_1=0$ and all its eigenvalues are real, since 
$\hbox{det}(\mathbb{G}-\lambda\mathbb{I})=
 \hbox{det}(\bar{\mathbb{G}}-\lambda\mathbb{I})$ where
$\bar{\mathbb{G}}$ is a symmetric matrix, namely
$\bar{\mathbb{G}}=
      \mathbb{H}\mathbb{K}\mathbb{L}\mathbb{K}\mathbb{H}$ 
with $\mathbb{L}$ being
the Laplacian of the network \cite{pecora98,motter04}, and matrices 
$\mathbb{H}$ and $\mathbb{K}$ being the diagonal matrices with elements
$H_{ii}={\cal N}_i^{-1/2}$ and $K_{ii}=k_i^{\alpha/2}$ respectively.

From Eq.~(\ref{linestab}), taking into account the ordering of the
eigenvalues $\lambda_i$, one easily concludes that the stability condition
for chaotic maps reads
\begin{equation}
\frac{1-\exp{(-\bar{\lambda})}}{\lambda_2}<\varepsilon<
\frac{1+\exp{(-\bar{\lambda})}}{\lambda_L},
\label{epsint}
\end{equation}
where $\bar{\lambda}$ is the Lyapunov exponent of the local  map.
In particular there is a range of coupling strengths enabling 
synchronizability if $\lambda_L/\lambda_2<
(1+\hbox{e}^{-\bar{\lambda}})/(1-\hbox{e}^{-\bar{\lambda}})$ holds.
Therefore, by computing the eigenvalues of the Laplacian matrix
one is able to find the range of couplings for which coherent states are
stable. For more detailed results see Ref.~\cite{new}.

Instead of starting from coherent states and studying
their stability to perturbations, in this paper we consider large samples 
of random 
initial configurations and study how much and under what conditions they
converge toward a coherent state.
This procedure not only reveals the existence of stable solutions but 
also gives a rough measure of its basin of attraction.

Earlier results \cite{jost01} show a transition to full
synchronization for two particular values of the nonlinearity $a$ 
in the homogeneous regime ($\alpha=0$), when either the coupling
strength or the number of outgoing connections are varied. 
Here, we show that the threshold value of such a transition as a
function of coupling strength and outgoing connectivity obeys a
power-law with an exponent that depends on the nonlinearity.
We study not only the usual {\it random} scale-free
network of Barab\'asi and Albert \cite{barabasi99}, but also 
{\it deterministic} scale-free networks constructed
in an iterative way \cite{barabasi01,dorogovtsev02,hanspriv}.
Deterministic scale-free networks are analytically easier to handle
\cite{barabasi01,iguchi04}.
Deterministic networks are applied for instance in spin systems
\cite{hanspriv}, and geographical and social networks
\cite{hanspriv,gonzalez04}.

We consider in Section \ref{sec:barabasi} the homogeneous
coupling regime for the random scale-free topology.
In Section \ref{sec:deter} we extend our results to two deterministic
scale-free networks, namely a pseudo-fractal network
\cite{dorogovtsev02} and an Apollonian network \cite{hanspriv}.
Discussion and conclusions are given in Section \ref{sec:discussion}.
\begin{figure}[htb]
\begin{center}
\includegraphics*[width=8.5cm]{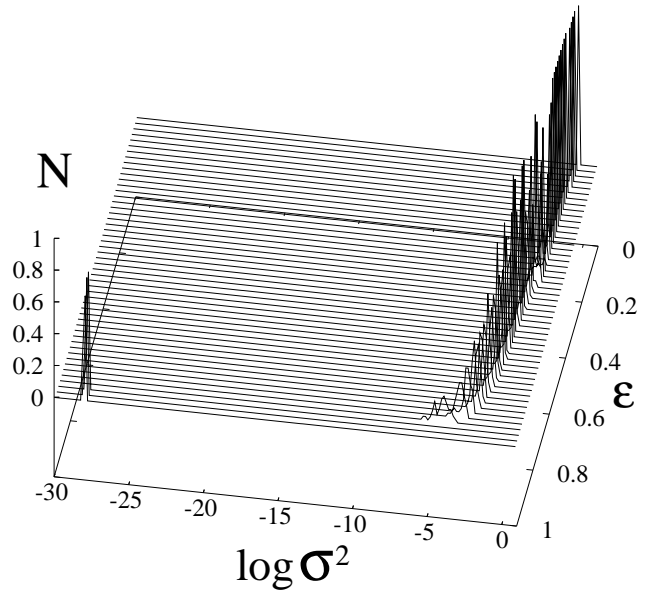}
\end{center}
\caption{\protect 
   Typical histogram of the standard mean square amplitude deviation
   $\sigma^2$ as a function of the coupling strength $\varepsilon$,
   showing the transition to coherence for a sample of
   $500$ initial random configurations.
   Here $N$ represents the fraction of configurations, and we discarded
   transients of $10^4$ time steps and fixed
   nonlinearity $a=2$, connectivity $k=m_0=8$, and number of nodes 
   $L=1000$, and $\alpha=0$. The base of the logarithm is $10$.}
\label{fig1}
\end{figure}
\begin{figure}[htb]
\begin{center}
\includegraphics*[width=8.5cm]{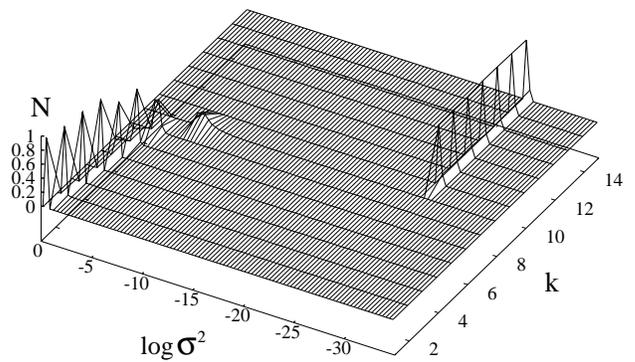}
\end{center}
\caption{\protect 
   Typical histogram of the standard mean square-deviation 
   $\hbox{log}_{10}\ \sigma^2$
   as a function of connectivity $k$, for $\varepsilon=0.95$, 
   $a=2$, $\alpha=0$ and $L=1000$. The same conditions and 
   initial configurations of Fig.~\ref{fig1} were used.}
\label{fig2}
\end{figure}

\section{Random scale-free networks}
\label{sec:barabasi}

Random scale-free networks share with many real networks, e.g.~the
World Wide Web, two generic mechanisms: growth and preferential
attachment \cite{albert02}. 
In this Section, we use the algorithm of Barab\'asi and
Albert \cite{albert02,barabasi99} to construct the network:
starting with a small number of nodes, say $m_0$, fully interconnected
with each other, one adds iteratively a new node with $k$ new
edges, which connect randomly the new node with previous
nodes, depending on their own number of connections.
As a general feature, one finds \cite{albert02} a connectivity
distribution which follows a power law with an exponent $\gamma=3$, 
independently on $m_0$ and $k$.
After a certain number of iterations, ones has a network with $L$
nodes, and then we place chaotic maps at the nodes, according to
Eq.~(\ref{model}), and observe if they synchronize or not after some
transient.

A suitable approach to study synchronization of chaotic
oscillators on an arbitrary network topology \cite{jost01} is to compute the
standard mean square-deviation 
\begin{equation}
\sigma^2_t=\tfrac{1}{L}\sum_{i=1}^L(x_{t,i}-\bar{x}_t)^2\; , 
\end{equation}
where $\bar{x}_t$ is the average amplitude at a 
given time step $t$. 
As one easily sees, all the nodes are synchronized at the same
amplitude whenever $\sigma^2$ is zero within numerical precision,
i.e.~$\sigma^2\sim 10^{-30}$.
We call these fully synchronized states {\it coherent states}, to
distinguish them from {\it partially} synchronized configurations,
when several different clusters of nodes with the same amplitude are
observed \cite{lind04}. 
\begin{figure}[t]
\begin{center}
\includegraphics*[width=8.2cm]{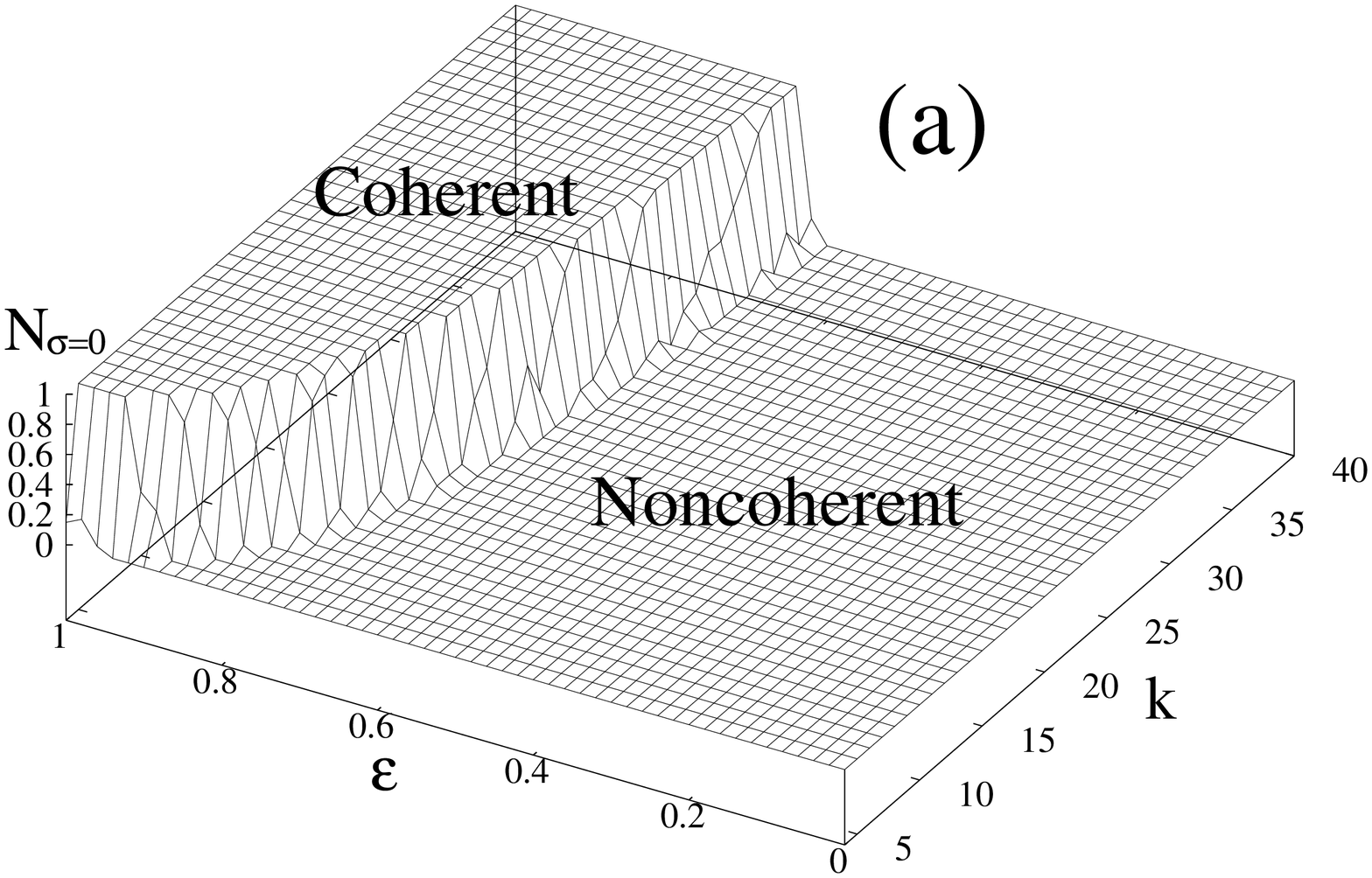}
\includegraphics*[width=8.2cm]{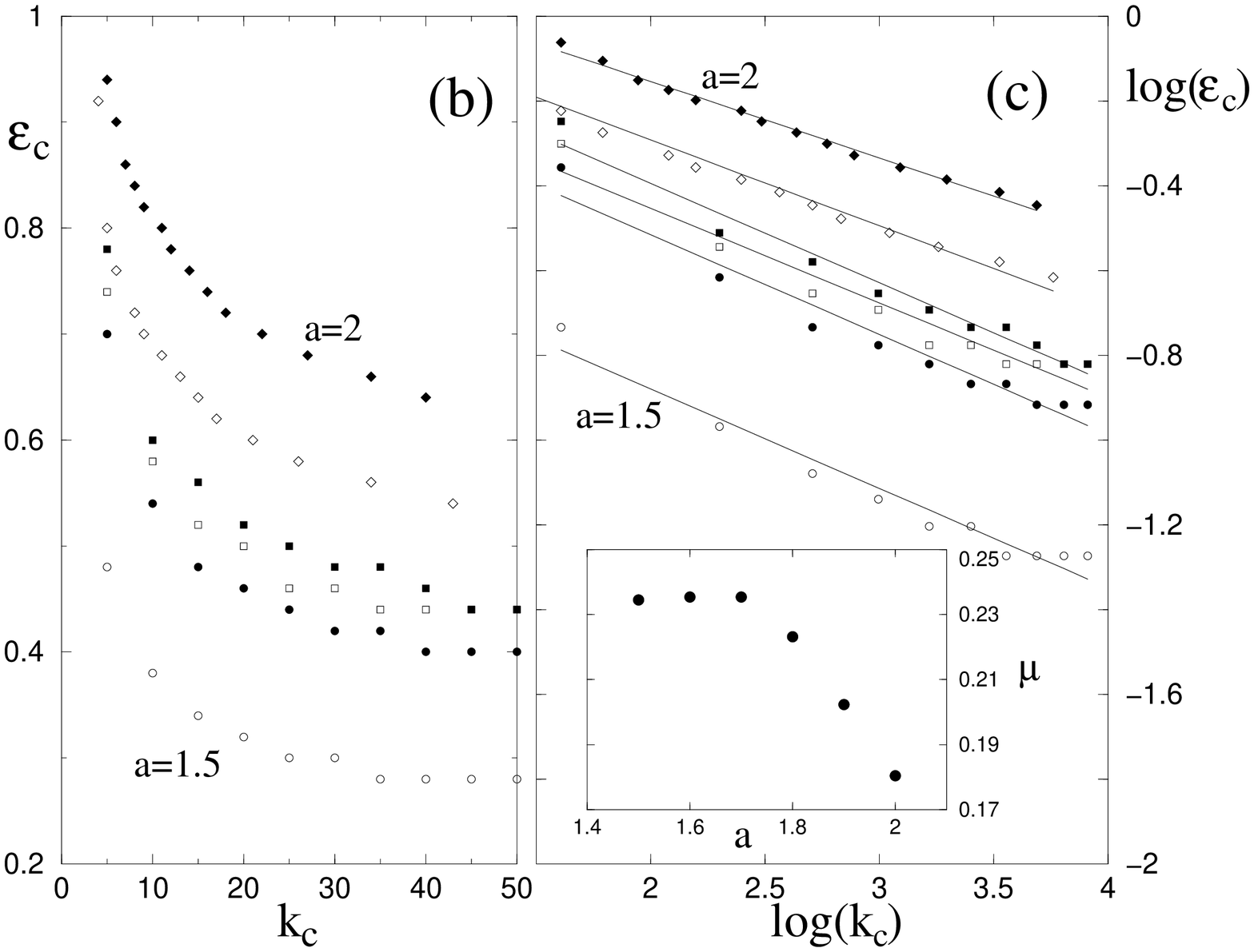}
\includegraphics*[width=8.2cm]{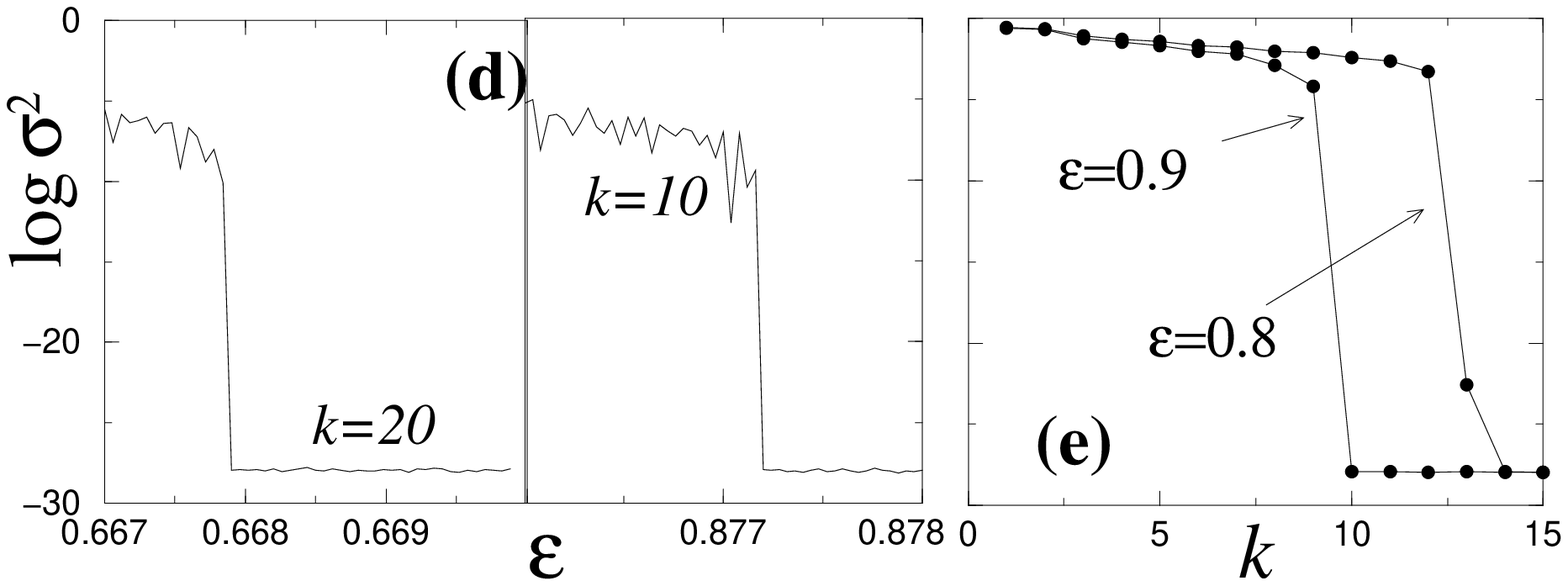}
\end{center}
\caption{\protect 
  Transition to coherence as a function of connectivity $k$ and coupling
  strength $\varepsilon$. 
  {\bf (a)} Fraction $N_{\sigma=0}$ of coherent states from 
  $500$ random initial configurations for $a=2$.
  {\bf (b)} Coherence transition curves in the $(\varepsilon,k)$ plane
  for 
  (from bottom to top) $a=1.5, 1.6, 1.8, 1.7, 1.9$ and $a=2$, and
  {\bf (c)} the same transition in a log-log plot (base $10$), showing
  power-law dependence between connectivity and coupling strength for
  the transition curves, with an exponent $\mu$ which depends on the
  value of $a$ (see inset).  
  Here $\alpha=0$, $L=1000$ and we used transients of $10^4$ time
  steps.
  By increasing the transient size to $\sim 10^6$ one sees
  clearly that the transition to coherence is of first-order 
  either {\bf (d)} when varying the coupling strength
  $\varepsilon$ or {\bf (e)} when varying the outgoing connectivity
  $k$.}
\label{fig3}
\end{figure}

In a previous work by Jost and Joy \cite{jost01}, concerning lattices
of coupled maps with different coupling topologies, a
transition to coherence between chaotic maps was found, when considering the
Barab\'asi-Albert network, occurring for particularly high coupling
strengths, typically of the order of $\varepsilon_c\sim 0.9$. 
Our simulations have shown that these transitions occur after
discarding transients of $\sim 10^4$ time steps and they do not change 
significantly with the network size.
Moreover, this transition to coherence is robust with respect to
initial configurations.

Figure \ref{fig1} shows a typical histogram of the standard mean
square-deviation as a function of the coupling strength $\varepsilon$, 
computed from a sample of 500 initial configurations, and fixing
$L=1000$, $a=2$, and $k=m_0=8$.
From the histogram, one clearly sees the sharp transition to coherence
and also its robustness to initial configurations, since for each
coupling strength all the final configurations have approximately the
same standard mean square-deviation.
In particular, above the threshold $\varepsilon_c\sim 0.9$, all 
initial configurations converge toward a coherent state, indicating
that in this parameter region the basin of attraction of coherent
states fills almost the entire phase space. 
These two features, sharp transition to coherence and robustness with
respect to initial configurations, are also observed when varying the
connectivity $k$, as illustrated by Fig.~\ref{fig2}.
Both Figs.~\ref{fig1} and \ref{fig2} were drawn fixing one of the
parameters, $\varepsilon$ or $k$. Our simulations
show that for the fully chaotic regime ($a=2$) the transition to
coherence occurs for gradually smaller coupling strength if the
connectivity $k$ is increased.
Figure \ref{fig3}a illustrates this fact,
plotting the fraction $N_{\sigma=0}$ of initial configurations which
converge to a coherent state. One sees a clear transition to coherence.
Computing similar histograms for other values of $a$, smaller then
$a=2$, and projecting them in the $(\varepsilon,k)$ plane one observes
similar transition lines in ranges with smaller coupling
strengths.
Figure \ref{fig3}b illustrates this fact by plotting the threshold
values, $\varepsilon_c$ and $k_c$, at the transition
curves for 
(from bottom to top) $a=1.5, 1.6, 1.8, 1.7, 1.9$ and $2$,
in the same conditions as in Fig.~\ref{fig3}a.
For all these values of $a$, the single uncoupled map shows chaotic
orbits, or at least the orbits have very large periods $\tau > 10^4$.
Note that the curve for $a=1.8$ is {\it below} 
that for $a=1.7$; this slight discrepancy is due to the fact that for $a=1.8$
the amplitudes of the logistic map vary (chaotically)
in a smaller interval than that observed for $a=1.7$.
As illustrated in Fig.~\ref{fig3}c, all curves obey, within our
statistical precision, a power-law, 
\begin{equation}
\varepsilon_c\propto k_c^{-\mu} .
\label{powerlaw}
\end{equation}
For the six values of $a$ above, the exponents are respectively
$\mu=0.2345, 0.2354, 0.2353, 0.2231, 0.2023$ and $0.1804$.
In other words, the exponent is almost constant below $a\sim 1.7$ and
decreases above this value, as illustrated in the inset of
Fig.~\ref{fig3}c.

In order to determine the nature of the transition to coherence seen
in Fig.~\ref{fig3}a, we show in Fig.~\ref{fig3}d a high-resolution
plot of $N_{\sigma=0}$ as a function of $\varepsilon$ for different
connectivities. 
Here one clearly sees a well-defined jump indicating that the
transition to coherence is of first-order.
One also observes first-order phase transitions when the outgoing
connectivity $k$ is varied (see Fig.~\ref{fig3}e).  
That transitions are indeed of first order is easily
recognized by the clear existence of  hysteresis:
when increasing either $\varepsilon$ or $k$ the
configuration eventually falls into a coherent state, no longer
spontaneously desynchronizing, no matter how far
the parameters are tuned back.

In this Section we only consider the case of homogeneous coupling
($\alpha=0$). For $\alpha>0$, when
the coupling to nodes with large number of neighbors is strengthened,
we find transitions to coherence similar to the ones illustrated in
Fig.~\ref{fig3}, occurring at weaker coupling strengths.

As a general conclusion one could say that, although the 
exponent $\gamma$ of the power-law distribution of connections
characterizing scale-free networks does not depend on the outgoing
connectivity $k$ \cite{barabasi99}, synchronization behavior is quite
sensitive to this quantity.

\section{Deterministic scale-free networks}
\label{sec:deter}

In the previous Section we focused on random scale-free networks,
i.e.~growing networks where new nodes are connected following
probabilistic rules.  
Although this stochasticity is typical for real networks, it is
more difficult to handle analytically \cite{barabasi01}.   
In this Section we study a different type of
networks: deterministic scale-free networks
\cite{barabasi01,dorogovtsev02,hanspriv}.

In particular, we use two different deterministic topologies, namely
the pseudo-fractal scale-free network introduced by Dorogovtsev et al
\cite{dorogovtsev02}, which is similar to the first deterministic
scale-free network proposed by Barab\'asi et al \cite{barabasi01}, and
was recently applied, e.g.~to studies of opinion formation
\cite{gonzalez04}, and the Apollonian network introduced by Andrade et
al \cite{hanspriv}.

\begin{figure}[t]
\begin{center}
\includegraphics*[width=4.2cm]{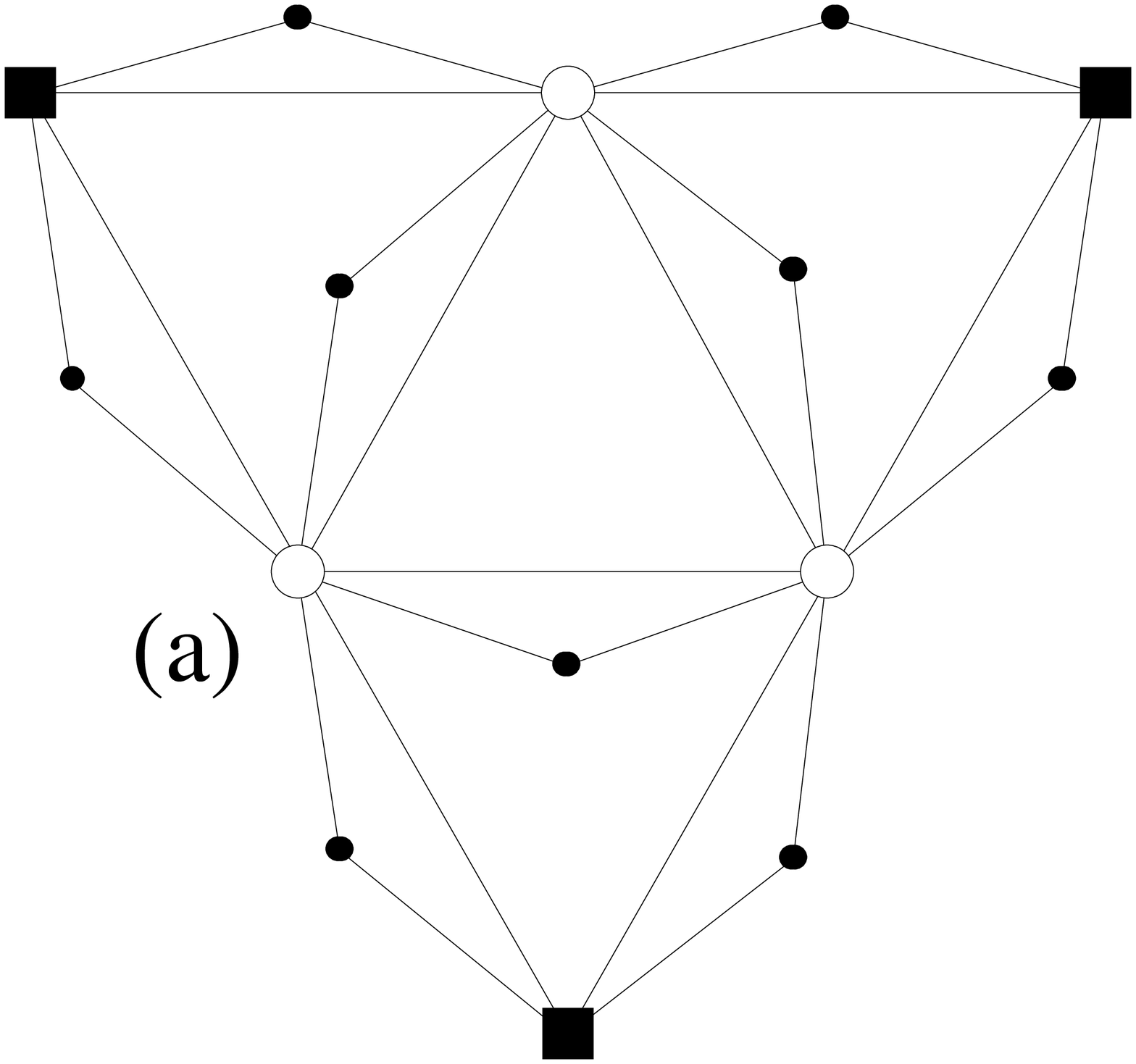}%
\includegraphics*[width=4.2cm]{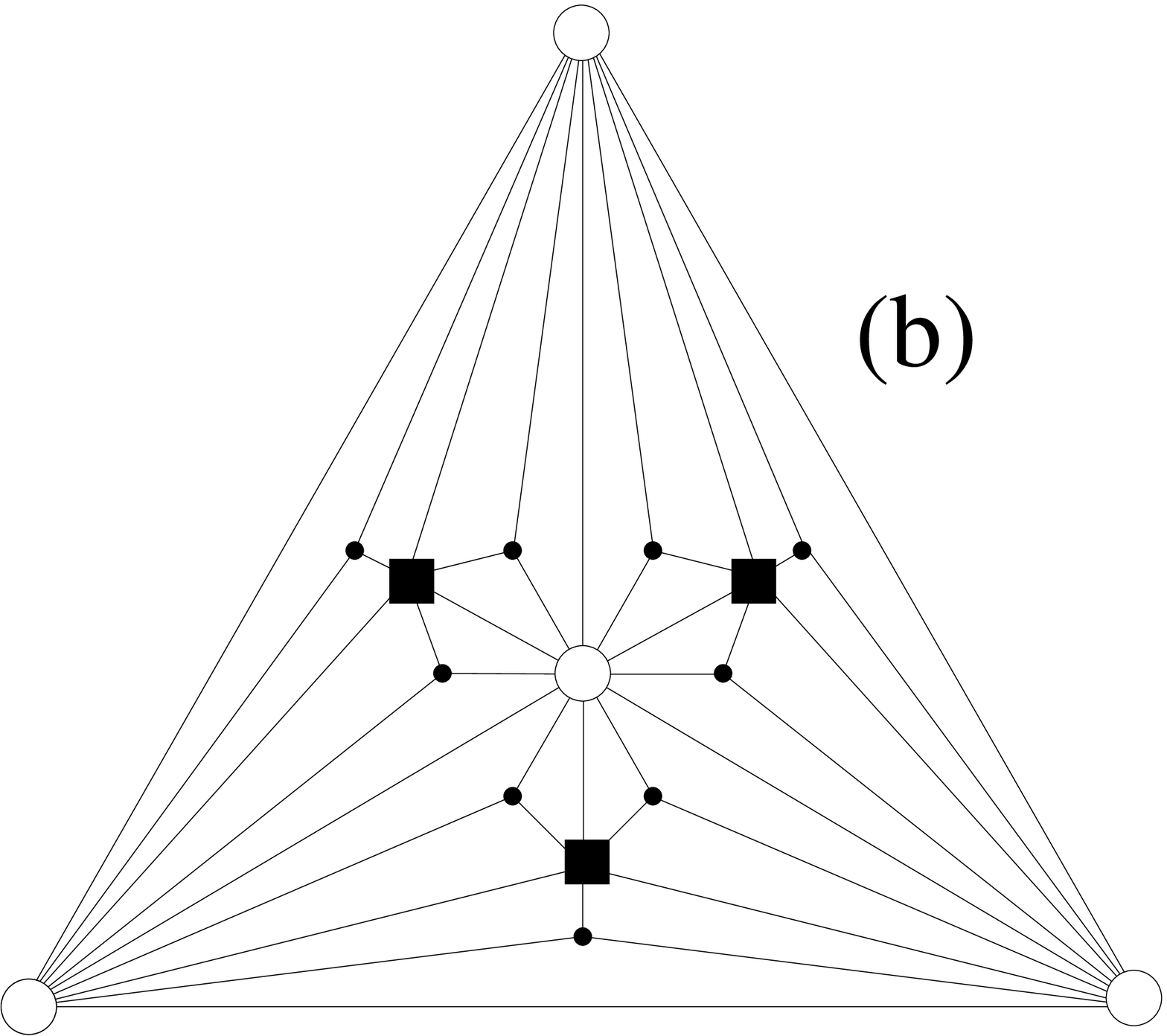}
\end{center}
\caption{\protect 
   Illustrations of two deterministic scale-free networks:
   {\bf (a)} the pseudo-fractal network \cite{gonzalez04}, and
   {\bf (b)} the Apollonian network \cite{hanspriv}.
   Identical symbols label nodes belonging to the same generation $n$
   (see text), namely $\bigcirc$ for $n=0$, $\blacksquare$ for $n=1$
   and $\bullet$ for $n=2$.} 
\label{fig4}
\end{figure}

The pseudo-fractal network of Dorogovtsev is obtained, starting from
three nodes interconnected with each other, and at each iteration
each edge generates a new node, attached to its two vertices.
Figure \ref{fig4}a illustrates this network after three iterations, i.e.~with
three generations of nodes. 
With such a construction the number of nodes $L_n$ and
the number of connections $M_n$ increases as
\cite{dorogovtsev02}
\begin{subequations}
\begin{eqnarray}
L_n &=& \tfrac{3}{2}(3^n+1)\; ,\label{pseudoL}\\
M_n &=& 3^{n+1}\; ,\label{pseudoM}
\end{eqnarray}
\end{subequations}
where $n$ is the number of iteration steps (generations). 
Moreover, at iteration $n$ the number of nodes with degree
$k=2,2^2,\dots,2^{n-1},2^n$ and $2^{n+1}$ is equal to
$3^n,3^{n-1},\dots,3^2,3$ and $3$ respectively,
yielding a power-law distribution with exponent
$\gamma=1+\ln{3}/\ln{2}\simeq 2.585$.

The Apollonian network has a construction algorithm different from the
pseudo-fractal network:
one starts with three interconnected nodes, defining a triangle.
At $n=0$ one puts a new node at the center of the triangle, joined to
the three other nodes, and thus defining three new smaller triangles.
At iteration $n=1$ one adds at the center of each of these three triangles
a new node, connected to the three vertices of the triangle, defining
nine new triangles; at iteration $n=2$ one adds one new node at the
center of each of these nine triangles, and so on (see
Fig.~\ref{fig4}b). 
With this construction procedure one obtains a deterministic
scale-free network \cite{hanspriv}, where the number of nodes $L_n$
and the number of connections $M_n$ are given respectively by
\begin{subequations}
\begin{eqnarray}
L_n &=& \tfrac{1}{2}(3^{n+1}+5)\; ,\label{apollonL}\\
M_n &=& \tfrac{3}{2}(3^{n+1}+1)\; .\label{apollonM}
\end{eqnarray}
\end{subequations}
At iteration $n$, the number of nodes with degree $k=3,3\cdot 2,3\cdot
2^2,\dots, 3\cdot 2^{n-1},3\cdot 2^{n}$ and $2^{n+1}$ is equal to
$3^n,3^{n-1},3^{n-2},\dots,3^2,3,1$ and $3$ respectively, yielding a
power-law distribution with the same exponent $\gamma$ as the one
found for the pseudo-fractal network.

From Fig.~\ref{fig4} and the description above, one easily concludes
that for the pseudo-fractal network the outgoing connectivity is fixed
at $k=2$, while for Apollonian networks one has $k=3$.
Despite both networks have a small number
of outgoing connections, they are quite different from the geometrical
point of view.
In fact, while the pseudo-fractal network has no metric, Apollonian
networks are embedded in Euclidean space and fill it densely as
$n\to\infty$, being particularly suited for describing geographical
situations \cite{hanspriv}.

As mentioned by Barab\'asi et al \cite{barabasi01}, a strong
advantage of deterministic networks is that it is often possible to
compute {\it analytically} their properties, for example the adjacency
matrix, whose eigenvalue spectrum characterizes the topology
\cite{albert02}.
A simple way to write the adjacency matrix of the pseudo-fractal
network is
\begin{equation}
{\cal A}_{n} =
\left [
\begin{array}{cc}
{\cal A}_{n-1}  & {\cal M}_{n-1} \\
{\cal M}_{n-1}^T  & \emptyset
\end{array}
\right ]_{L_n \times L_n}  \; ,
\label{pseudo_adjmat}
\end{equation}
where $L_n$ is given by Eq.~(\ref{pseudoL}), ${\cal M}^T$ represents
the transpose matrix of ${\cal M}$ and for each generation
$n=1,2,\dots$ the matrix ${\cal M}_n$ reads
\begin{equation}
{\cal M}_n =
\left [
\begin{array}{ccc}
{\cal M}_{n-1} & {\cal M}_{n-1} & \emptyset \\
\emptyset      & \emptyset      & {\cal B}_{n-1}
\end{array}
\right ]_{2\cdot3^{n-1}\times 3^n}  \; ,
\end{equation}
with
\begin{equation}
{\cal B}_{n-1} =
\left [
\begin{array}{cccc}
{\cal A}_0 & \emptyset  &  \dots  & \emptyset \\
\emptyset  & {\cal A}_0 &  \dots  & \emptyset \\
\vdots     & \vdots      &  \ddots & \vdots    \\
\emptyset  & \emptyset  & \dots   & {\cal A}_0
\end{array}
\right ]_{3^{n-1}\times 3^{n-1}}   
\end{equation}
and whose starting form is
\begin{equation}
{\cal M}_0 = {\cal A}_0 =
\left [
\begin{array}{ccc}
0 & 1 & 1 \\
1 & 0 & 1 \\
1 & 1 & 0
\end{array}
\right ]_{3\times 3}  \; .
\end{equation}

For the Apollonian network, the adjacency matrix is given by the same 
recurrence of Eq.~(\ref{pseudo_adjmat}), but this time with
\begin{equation}
{\cal A}_0 =
\left [
\begin{array}{cccc}
0 & 1 & 1 & 1\\ 
1 & 0 & 1 & 1\\ 
1 & 1 & 0 & 1\\ 
1 & 1 & 1 & 0
\end{array}
\right ]\; ,
\label{a0_apollon_adjmat}
\end{equation}
and ${\cal M}_n$ being a matrix with $(3^n+5)/2$ rows and $3^n$ columns
and having in each column three non-zero elements only.

\begin{figure}[b]
\begin{center}
\includegraphics*[width=8.5cm]{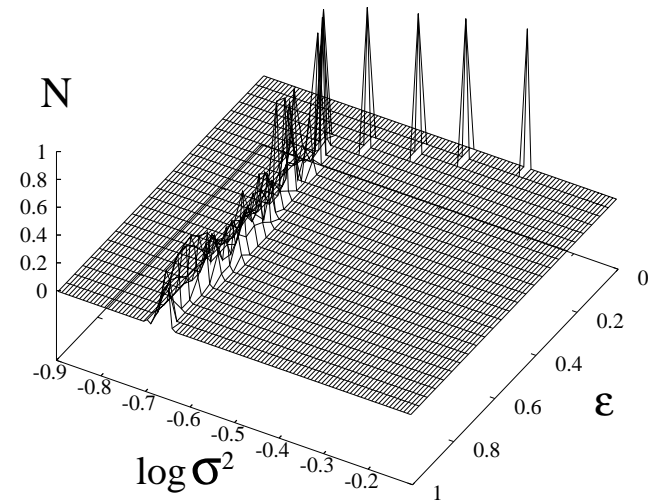}
\end{center}
\caption{\protect 
   Typical histogram of the standard mean square-deviation 
   $\hbox{log}_{10}\ \sigma^2$ for
   the pseudo-fractal network as a function of the coupling strength
   $\varepsilon$.  
   Similar result is obtained for the Apollonian network.
   Nodes are ruled by the map $f(x)=1-2x^2$ and we fixed $\alpha=0$.
   For the pseudo-fractal network we fixed the number of nodes
   $L=1095$ ($6$ generations of nodes), while for the Apollonian
   network $L=1096$ ($6$ generations of nodes). 
   For each coupling strength, the value $N$ indicates the fraction of
   $500$ initial configurations, and we used transients of $10^4$
   time steps.}
\label{fig5}
\end{figure}

For stability analysis purposes (see Section \ref{sec:intro}), 
one could derive the Laplacian matrices directly from these adjacency 
matrices, multiplying the adjacency matrices by $-1$ and adding the 
appropriate number of connections of each node $i$ along the main 
diagonal.

As shown in Fig.~\ref{fig5}, despite having quite similar structural
properties \cite{dorogovtsev02,hanspriv}, the global dynamics of the
entirely deterministic scale-free networks shows quite different
behavior than the one observed for the Barab\'asi-Albert model in the
previous section.
Namely, there is {\it no} coherence observed for the fully chaotic map
for $a=2$. 
In fact, from Fig.~\ref{fig5} one sees that the standard mean
square-deviation never vanishes.
Instead, it is characterized by some large value which is almost
constant beyond the weak coupling regime ($\varepsilon\gtrsim 0.2$).
In the weak coupling regime ($\varepsilon\lesssim 0.2$) the standard
mean square-deviation is even larger, since the coupling is not strong
enough to compensate the highly chaotic local dynamics ($a=2$).
Our simulations have shown that this feature remains valid for any
transient up to $10^6$ time steps, and it seems to be valid for any 
value of $a$ for which the quadratic map supports chaotic orbits. 
One possible physical explanation for this absence of synchronizability is
that long range random connections are crucial to improve the ability
for synchronization and, due to the deterministic construction of the 
network, there are no long range connections as in the Barab\'asi-Albert 
scale-free network.
Rigorously speaking, the absence of synchronizability is valid only within
the range $0\le\varepsilon\le 1$. 
Of course that, as long as condition (\ref{epsint}) holds, there is for sure
some range of coupling strengths for which the coherent states are stable.
However, since we are working with maps of the interval,
we neglect coupling strengths outside the unit interval, otherwise it is
not possible to guarantee convergence for {\it all} initial configurations.

\begin{figure*}
\begin{center}
\includegraphics*[width=9.0cm]{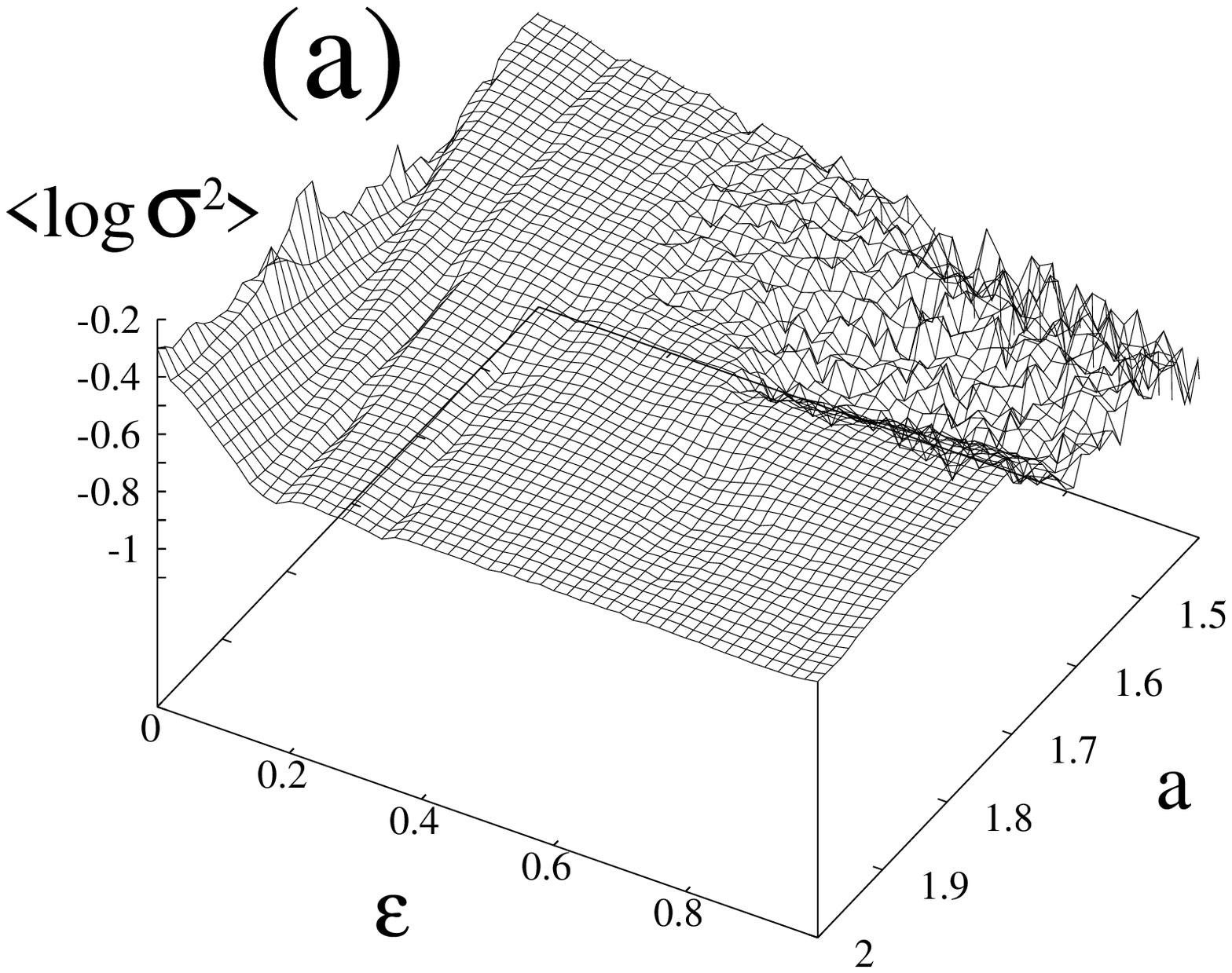}%
\includegraphics*[width=9.0cm]{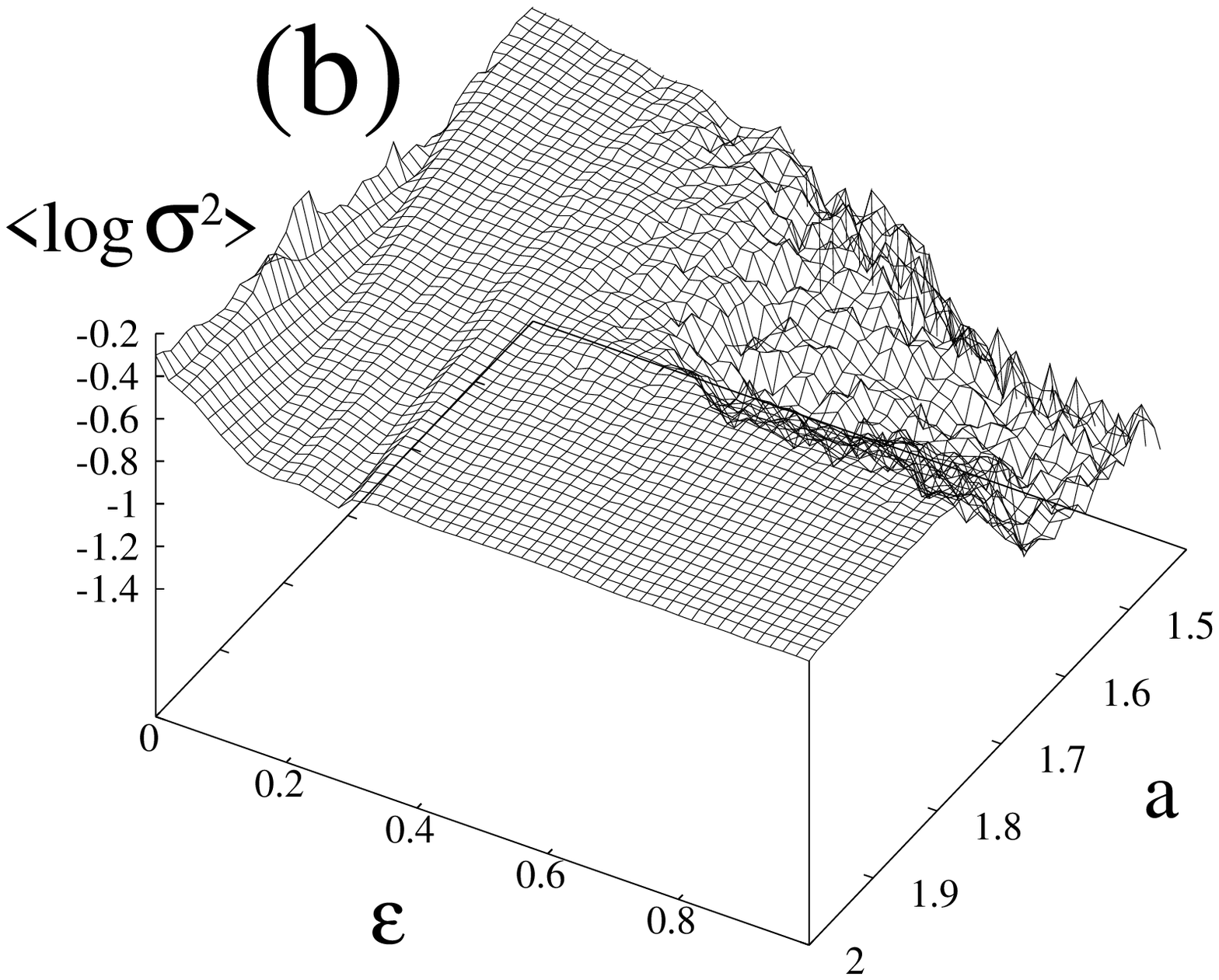}
\end{center}
\caption{\protect 
  Histogram of the standard mean square-deviation $\sigma^2$ as a
  function of nonlinearity $a$ and coupling strength $\varepsilon$,
  for deterministic scale-free networks, namely
  {\bf (a)} pseudo-fractal network and
  {\bf (b)} Apollonian network. 
  The mean square-deviation is averaged over a sample of $500$ initial
  configurations and during $100$ time steps, after discarding
  transients of $10^4$ time steps. 
  Here $\alpha=0$ and the base of the logarithm is $10$.}
\label{fig6}
\end{figure*}
\begin{figure*}[htb]
\begin{center}
\includegraphics*[width=18.0cm]{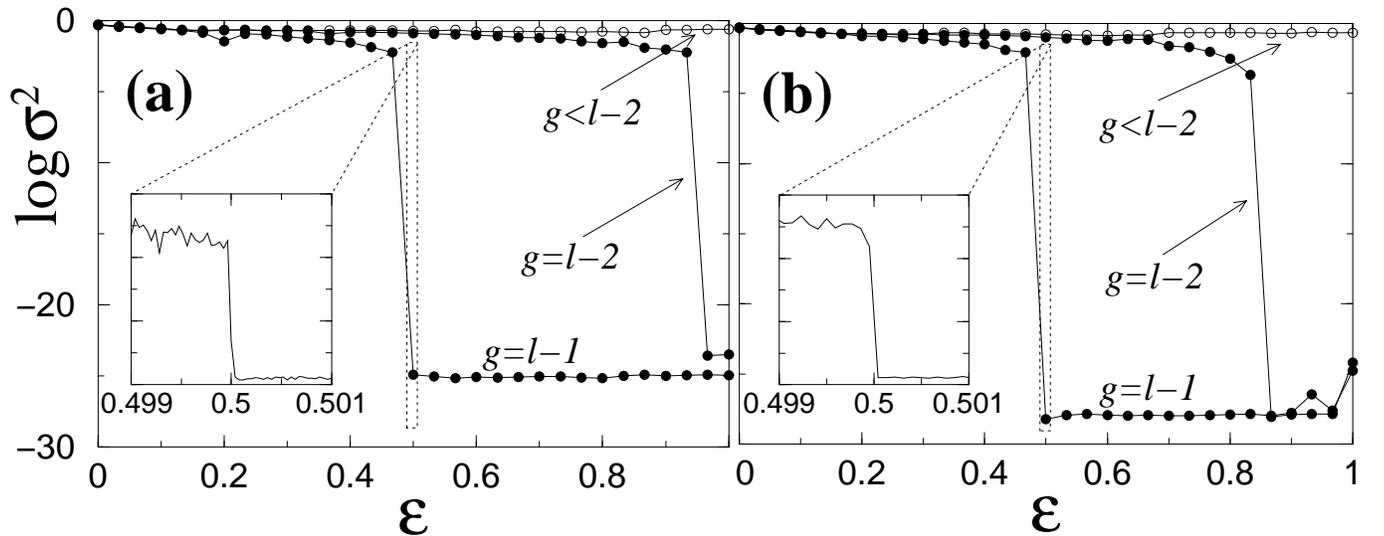}
\end{center}
\caption{\protect 
  Transitions to coherence in deterministic scale-free
  networks, when synchronizing the first $g$ generations of nodes out
  of $\ell$ generations (see text).
  {\bf (a)} pseudo-fractal network and
  {\bf (b)} Apollonian network. 
  The collective dynamical behavior is quite insensitive to hubs (see text).
  Insets show that transitions to coherence are of first-order.
  For each network, we use $\ell = 9$ generations of nodes and $a=2$ fixed.
  The base of the logarithm is $10$.} 
\label{fig7}
\end{figure*}
\begin{figure}[htb]
\begin{center}
\includegraphics*[width=8.5cm]{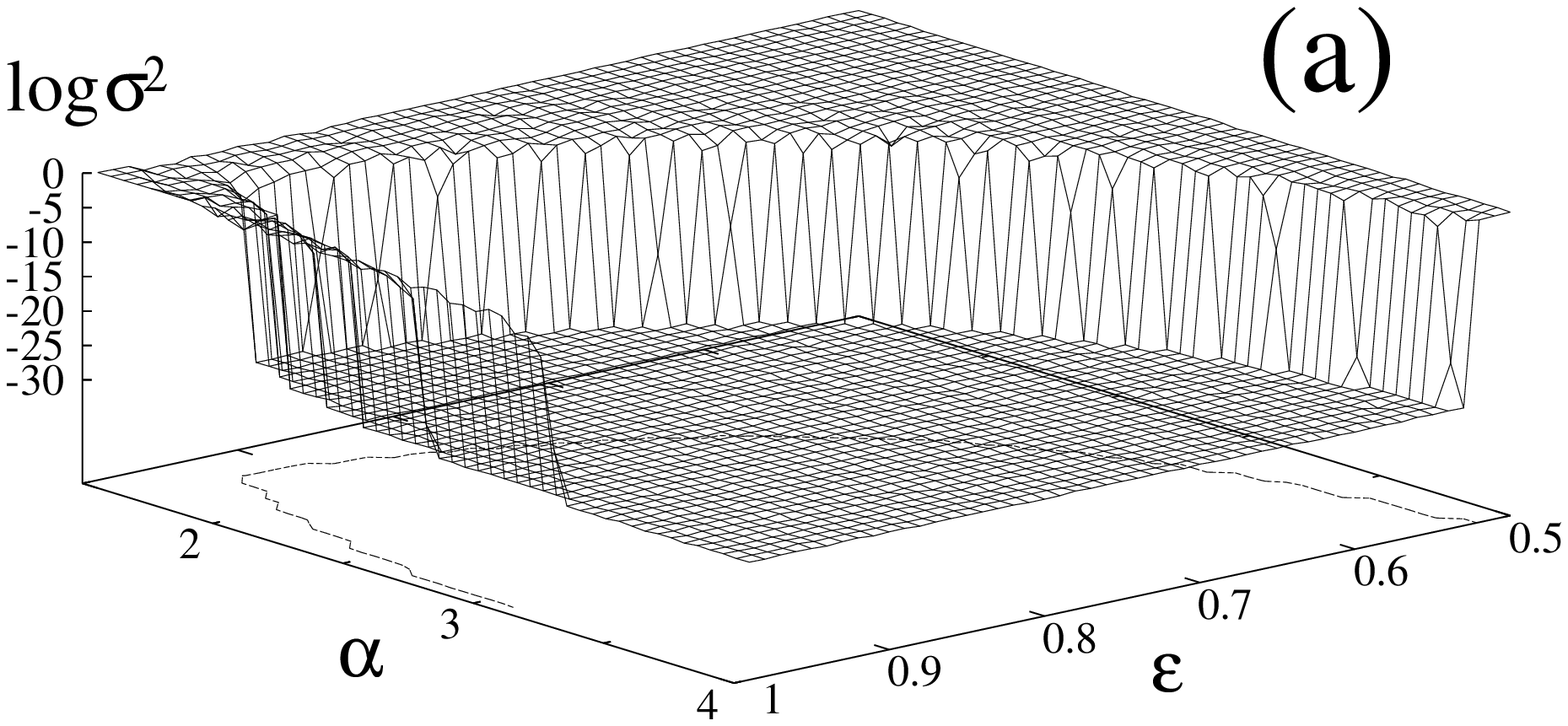}
\includegraphics*[width=8.5cm]{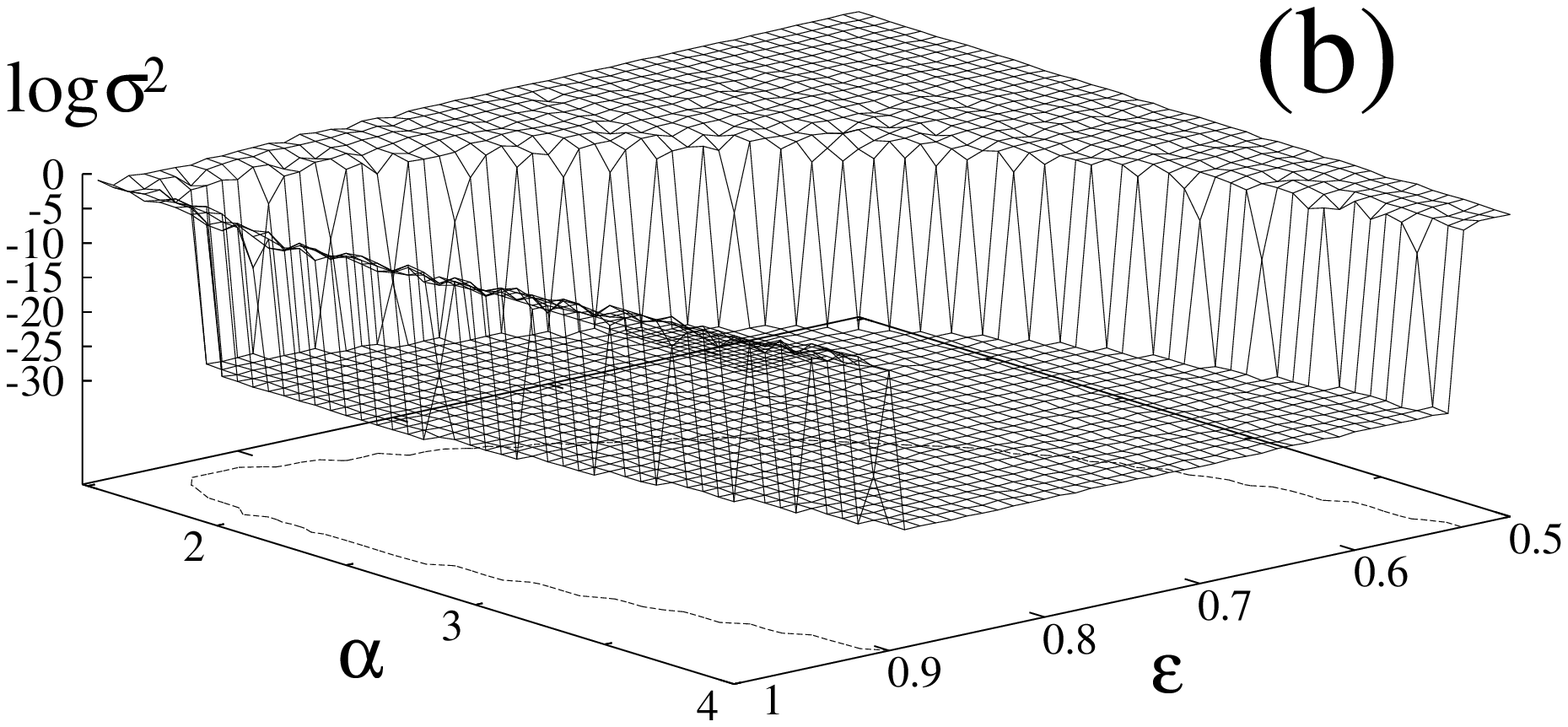}
\includegraphics*[width=8.5cm]{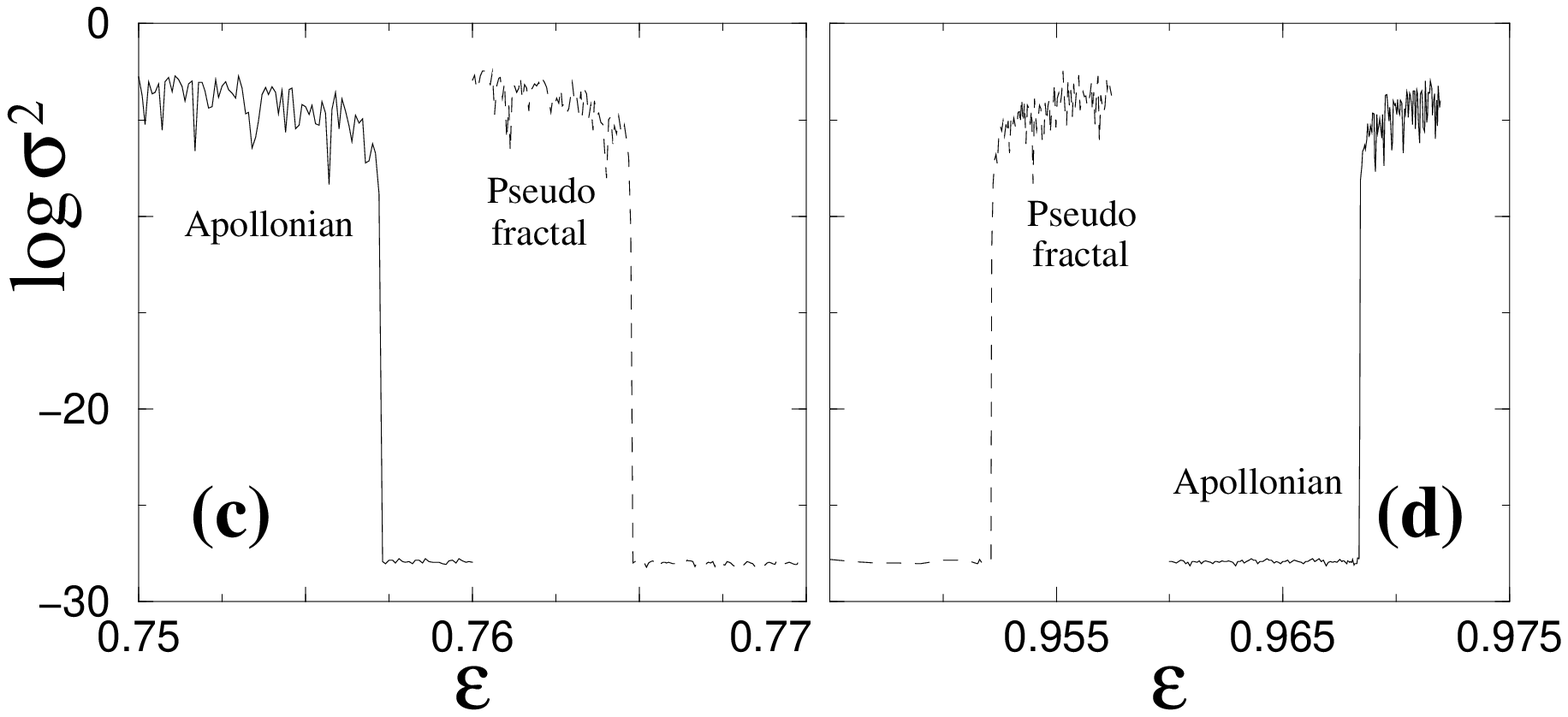}
\end{center}
\caption{\protect 
  Inducing transition to coherence by varying the heterogeneity
  $\alpha$ (see Eq.~(\ref{model})) in scale-free networks.
  {\bf (a)} pseudo-fractal network and
  {\bf (b)} Apollonian network. 
  For strong heterogeneity coherence appears beyond
  a relatively high coupling strength, but it
  disappears again for very large couplings (see text).
  For each network, we use $\ell = 6$ generations of nodes and fix
  $a=2$.
  {\bf (c)} and {\bf (d)} show high-resolution plots of $\sigma^2$ as
  a function of $\varepsilon$ for $\alpha=2$, emphasizing the
  first-order phase transition to coherence.
  The base of the logarithm is $10$.} 
\label{fig8}
\end{figure}

We plot the average $\langle \hbox{log}_{10}\ \sigma^2\rangle$ 
as a function of both the nonlinearity $a$ and the coupling strength 
$\varepsilon$ simultaneously. 
Figure \ref{fig6} shows this dependence for the full range of the
coupling strength $0\le\varepsilon\le 1$ and for nonlinearities
above the accumulation point of the first period-doubling cascade
of the quadratic map, namely $a\simeq 1.411$.
In Fig.~\ref{fig6}a we use a pseudo-fractal network, while
in Fig.~\ref{fig6}b the Apollonian network is considered.
In both cases six generations of nodes are taken, yielding a total
of $L=1095$ nodes for the pseudo-fractal network and $L=1096$
for the Apollonian network.

As can be seen from these figures, in both cases one has two main
regions: 
(I) a region where the standard mean square-deviation is large and
varies smoothly with the parameters and 
(II) a region where the mean square-deviation is smaller but has
larger fluctuations. 
For the Apollonian network the irregular region is characterized by 
significantly smaller values for the standard mean square-deviation.

The results observed in the histograms of Fig.~\ref{fig6} are somehow
surprising, since irregular variations of the standard mean
square-deviation occur for low nonlinearity and high coupling
strengths, precisely where one would expect the most regular
behavior of the node dynamics.
However, this irregularity is just apparent, since $\langle
\hbox{log}_{10}\ \sigma^2\rangle$ is an average over a sample of initial
conditions.
Whenever some initial configuration leads to coherence
the zero standard deviation decreases this average.
Therefore, for $a<1.7$, i.e.~in the region of irregular variations of
$\sigma^2$, coherent solutions are observed.
In fact, from the stability condition in Eq.~(\ref{epsint})
  one sees that for periodic maps, the lower boundary of the
  $\varepsilon$-range is always
  negative while the upper is positive, yielding always a finite
  range of coupling strengths where synchronizability is possible.
Since for Apollonian networks, the fluctuations occur at small values
of the standard deviation, this means that there is probably a larger
number of coherent solutions.

Figure \ref{fig6} indicates that there is a lack of coherent solutions
above $a\sim 1.7$.
To explain this fact one should notice that both the pseudo-fractal
and Apollonian networks have small outgoing connectivities, $k=2$ and
$k=3$ respectively.
Since one also observes almost no coherent solution for
random scale-free networks neither for $k=2$ nor for $k=3$ (see
Section \ref{sec:barabasi}), we believe that the outgoing 
connectivity $k$ is the main parameter
controlling synchronization between oscillators in complex networks.
By choosing another deterministic scale-free network
with a higher outgoing connectivity, say $k=10$, one might see
coherent solutions beyond a coupling threshold value approximately
similar to those computed for random scale-free networks (see
Fig.~\ref{fig3}).

In the remainder of this section we will consider the two deterministic
scale-free topologies, and study possible ways of inducing coherent
states.

Starting from a total number of $\ell$ generations, 
one efficient way of inducing coherence is by imposing synchronization
among a certain number of $g<\ell$ generations.
By generation we mean the set of new nodes appearing
simultaneously at a given iteration $n$, during the `construction' of the
network.
For instance, in the Apollonian network, the first generation has
$L_1=3$ nodes, the second has $L_2-L_1=9$ and the $n$th generation has
$L_{n}-L_{n-1}=3^{n}$ nodes.
In other words, we are interested in the collective effects when 
the most connected nodes (hubs) are synchronized.
We will show that, in fact, hubs play no dominant role for the
synchronization of nodes.

Figure \ref{fig7} shows the standard mean square-deviation as a
function of coupling strength for pseudo-fractal (Fig.~\ref{fig7}a)
and Apollonian networks (Fig.~\ref{fig7}b).
In each case we choose the fully chaotic map ($a=2$) and impose 
synchronization among the nodes of the first $g$ generations
by setting them to be their mean amplitude at each time-step.

For both networks, one sees from Fig.~\ref{fig7} that the standard
mean square-deviation remains large when synchronization is imposed to
all $g<\ell-2$ generations.
Coherent solutions are only observed for $g=\ell-2$ and $g=\ell-1$,
beyond a coupling threshold which is smaller for the latter
case. 
Surprisingly, for $g=\ell -1$ the transition to coherence occurs
precisely for the same coupling strength in both networks.
This may be due to the fact that, the fraction $L_g/L_{\ell}$ of nodes 
on which one imposes synchronization is approximately the same for
both networks.
For $g=\ell-2$ the pseudo-fractal network shows coherence only above
very high coupling strengths, near $\varepsilon\sim 1$, while for
Apollonian networks the threshold is much lower.
Although in this case the fraction of nodes on which one imposes
synchronization is also similar for both networks, it is much
smaller than in the case where $g=\ell -1$.
The transition to coherence occurs at different coupling
strengths because the number of synchronized nodes is not
enough to suppress the effect of the outgoing connectivity. So, since
the outgoing connectivity is larger for the Apollonian network, its
transition to coherence occurs for weaker coupling strengths.
For both networks, one obtains similar results for any higher 
value $\ell$ of generations  since the quotient of the number
of nodes between two successive generations $L_n/L_{n-1}\to 3$ as $n$ 
increases.

As a general remark, one observes from Fig.~\ref{fig7} that one needs
to synchronize a rather high number of generations to induce
coherence. 
Therefore, it seems that, dynamical collective behavior on scale-free
networks is quite insensitive to hubs. 
As shown in the insets of Fig.~\ref{fig7}a and
\ref{fig7}b, the transition to coherence is of first-order.

Another way to induce coherence in these two deterministic
scale-free networks is, instead of imposing synchronization to the
most connected nodes, to strengthen their coupling to the other nodes
by taking $\alpha> 0$ in Eq.~(\ref{model}). 
Figure \ref{fig8} illustrates the transition to coherence by
varying the heterogeneity $\alpha$ for the pseudo-fractal
(Fig.~\ref{fig8}a) and the Apollonian network (Fig.~\ref{fig8}b).
For both networks, one sees that coherence sets in for $\alpha\gtrsim
1.5$, and only beyond a certain threshold of the coupling strength.

In particular, one observes the remarkable
fact that coherence appears only in an intermediate coupling range,
i.e.~neither to large nor to small values.
This is in agreement with previous works \cite{pecora97} concerning other
systems of coupled chaotic oscillators, where one also observes that
synchronized chaos requires that the coupling must be neither too weak
nor too strong in order to avoid triggering spatial instabilities
\cite{strogatz01}.  
From Figs.~\ref{fig8}c and \ref{fig8}d one observes that all these
transitions to coherence are of first-order.

For the pseudo-fractal, the upper threshold disappears when the
heterogeneity $\alpha$ is further increased. 
However, for the Apollonian network the upper threshold not only
persists but it shifts toward smaller and smaller coupling strengths
when $\alpha$ is further increased.
As far as we know, this is the first time that one observes such
behavior, and it should be related to the geometrical differences
between both networks. In other words, since Apollonian is a very
particular scale-free network, being the only one, studied so far which
is embedded in Euclidean space, this particular feature seems to
enable nontrivial synchronization behavior: stronger dominance in the
coupling to the most connected nodes {\it destroys} coherence.

\section{Discussion and conclusions}
\label{sec:discussion}

In this paper we studied fully synchronized solutions for three
scale-free network topologies.
The main conclusion is the following:
in random scale-free networks synchronization of chaotic maps not only
depends on the coupling strength but is mainly controlled by the
outgoing connectivity $k$, which is a measure of cohesion in the networks.
Because of that, one finds coherent solutions in random scale-free
networks of fully chaotic logistic maps ($a=2$) with outgoing
connectivity $k=8$ and homogeneous coupling, but not in deterministic
scale-free networks, since they have rather small effective outgoing 
connectivity, namely $k=2$ for the pseudo-fractal network and $k=3$
for the Apollonian network.

Therefore, although the exponent $\gamma$ of connection distributions
in scale-free networks does not depend on the outgoing connectivity 
\cite{albert02}, we have shown that, in general, synchronization of
chaotic maps in such coupling topologies is quite sensitive to it. 
Moreover, the transition to coherence is of first-order, indicating a
similarity with other complex networks \cite{strogatz01}.
In particular, the threshold values of the coupling strength obey a
power-law, Eq.~(\ref{powerlaw}), as function of the outgoing
connectivity. 
The exponent of this power-law depends on the nonlinearity $a$ of the
chaotic map, being almost constant below $a_c\sim 1.7$ and decreasing
linearly above it. Interestingly this value of $a_c$ is in the
vicinity of the bifurcation of the quadratic map where the period-$3$
window appears, and coincides with the appearance of other
nontrivial behaviors in coupled map lattices with regular topologies,
namely in the velocity distribution of traveling wave solutions
\cite{lind04b}.

The synchronization criterion was based here in the square mean standard
deviation following previous studies \cite{jost01}.
Of course,  it could be possible to have numerically a zero
standard deviation $\sigma^2\sim 10^{-30}$ with a particular
oscillator slightly nonsynchronized.
However, in such case the deviation of the oscillator amplitude 
from the rest of the network would be of the order of
$10^{-15}$, a majorant of the precision for our criterion.
Therefore, we believe that within this numerical precision there are
no spurious results.
Further investigations could be done, implementing  extensions of
clustering criteria such as, e.g., that of 
Pikovsky et al.~\cite{pikovsky01}.

For deterministic scale-free networks with homogeneous coupling, the
same value $a_c$ indicates the threshold above which no coherent
solutions are observed, independently of the coupling strength.
Above $a_c$, coherence is observed only for heterogeneous
coupling, namely for $\alpha\gtrsim 1.5$.
However, for this range of values, we have also shown that coherence is  
also absent either for very small or for very large coupling
strengths, due to spatial instabilities.
Another particularly interesting result that still needs to be
explained is that, for Apollonian networks, the coupling threshold
beyond which coherence disappears gets smaller when the
heterogeneity is further increased. 
This point is not observed for the pseudo-fractal network and may be
due to the geometrical differences between both deterministic networks.

As a general property, we have shown that all transitions to
coherence are of first-order.
Furthermore, all results are robust not only against changes of the
initial configurations of node amplitude but also, in random
scale-free networks, against changes of the connection network. 
We also presented preliminary results indicating that in
scale-free networks hubs play apparently no fundamental role in the
dynamical collective behavior, what remains to be further investigated.

\section*{Acknowledgments}

The authors thank J.S.~Andrade Jr., A.O.~Sousa, M.C.~Gonz\'alez,
for useful discussions. 
P.G.L.~thanks Funda\c{c}\~ao para a Ci\^encia e a Tecnologia, Portugal, for
financial support. 
J.A.C.G.~thanks Conselho Nacional de Desenvolvimento Cient\'{\i}fico e
Tecnol\'ogico, Brazil  and Sonderforschungsbereich 404, 
Germany, for financial support.



\end{document}